\documentstyle[12pt]{ioplppt}          

\newcommand{\be}{\begin{equation}}
\newcommand{\ee}{\end{equation}}
\newcommand{\ba}{\begin{eqnarray}}
\newcommand{\ea}{\end{eqnarray}}
\newcommand{\brr}{\begin{array}}
\newcommand{\err}{\end{array}}
\newcommand{\bc}{\begin{center}}
\newcommand{\ec}{\end{center}}
\begin{document}
\title{Relativistic Cosmology: from super--horizon to small scales} 
\author{Sabino Matarrese} 
\address{Dipartimento di Fisica `Galileo Galilei', Universit\`a di Padova \\ 
via Marzolo 8, I--35131 Padova, Italy }

\begin{abstract}

The general relativistic non--linear dynamics of a self--gravitating 
collisionless fluid with vanishing  vorticity is studied in synchronous and 
comoving -- i.e. {\em Lagrangian} -- coordinates. Writing the equations in 
terms of the metric tensor of the spatial sections orthogonal to the fluid 
flow allows an unambiguous expansion in inverse powers of the speed of light. 
The Newtonian and post--Newtonian approximations are derived in Lagrangian 
form; the non--linear evolution of the system on super--horizon scales, leading 
to the so--called ``silent universe", is also briefly discussed. A general 
formula for 
the gravitational waves generated by the non--linear evolution of cosmological 
perturbations is given: a stochastic gravitational--wave background is 
shown to be produced by non--linear cosmic structures, with present--day 
closure density $\Omega_{gw} \sim 10^{-5}$ -- $10^{-6}$ on the scale of 
1 -- 10 Mpc. 

\end{abstract} 

\section{Introduction}

In these notes I review some of the material which I presented in my 
lectures. These actually covered two different topics: the first was 
inflation in the early universe, mostly from a kinematical point of view, with 
some emphasis on the evolution of irregularities outside the Hubble radius, 
while the second was 
the dynamics of self--gravitating irrotational dust in Lagrangian coordinates. 
A strong link between the two topics is given by the use of General 
Relativity in dealing with cosmological problems on 
both very large (super--horizon) and small scales, where it is usually 
thought that Newtonian gravity should provide a detailed description of 
the system. Given the many excellent reviews on cosmological inflation, among 
which the lectures given by Rocky Kolb at this school (this volume), I 
preferred to devote these notes to the second topic. Most of the material 
presented here is based on a recent work in collaboration with David Terranova 
\cite{bi:pn}, as well as on many ideas which come out from 
a research program on ``silent universes", in collaboration with Ornella 
Pantano, Diego Saez and Marco Bruni \cite{bi:mps93},\cite{bi:mpsprl},
\cite{bi:mps94},\cite{bi:silent},\cite{bi:nohair}. The origin of my interest, 
however, goes back to my first reading of the lectures by George Ellis at the 
1969 Varenna School \cite{bi:ellisvar}, whose title, ``Relativistic Cosmology", 
I borrowed for these notes. 
Let me finally mention that the material in these notes is strongly related to 
that in the lectures by Francois Bouchet and Thomas Buchert (this volume), 
which deal with various aspects of the dynamics of self--gravitating dust 
within the Newtonian approximation. \\

The first aim of these notes is to try to convince the skeptical reader 
that a general relativistic approach is relevant for the structure 
formation problem in cosmology, i.e. that there exists a class of problems 
in which non--linearity and relativistic effects both come into play. 
The gravitational instability of cold matter 
in a cosmological framework is in fact usually studied within the Newtonian 
approximation. This basically consists in neglecting terms higher than the 
first in 
metric perturbations around a matter--dominated Friedmann--Robertson--Walker 
(FRW) background, while keeping non--linear density and velocity perturbations. 
This approximation is usually thought to produce accurate 
results in a wide spectrum of cosmological scales, namely on 
scales much larger than the Schwarzschild radius of collapsing bodies and 
much smaller than the Hubble horizon scale, where the peculiar gravitational 
potential $\varphi_g$, divided by the square of the speed of light $c^2$ to 
obtain a dimensionless quantity, keeps much less than unity, while the peculiar 
matter flow never becomes relativistic. 
To be more specific, the Newtonian approximation 
consists in perturbing only the time--time component of the FRW 
metric tensor by an amount $2\varphi_g/c^2$, where $\varphi_g$ 
is related to the matter density fluctuation $\delta$ via the 
cosmological Poisson equation, 
$\nabla_x^2 \varphi_g ({\bf x},t) = 4 
\pi G a^2(t) \varrho_b(t) \delta({\bf x}, t)$, 
where $\varrho_b$ is the background matter density and 
$a(t)$ the appropriate FRW scale--factor; 
the Laplacian operator $\nabla_x^2$ has been used here with its standard 
meaning of Euclidean space. The fluid dynamics is then 
usually studied in Eulerian coordinates by accounting for mass conservation 
and using the cosmological version of the Euler equation for a 
self--gravitating pressureless fluid, as long as the 
flow is in the laminar regime, to close the system. 
To motivate the use of this ``hybrid approximation", which 
deals with perturbations of the matter and the geometry at a different 
perturbative order, one can either formally expand the correct 
equations of GR in inverse powers of the speed of light
(e.g. Ref.\cite{bi:weinberg}), or simply notice that the peculiar 
gravitational potential is strongly suppressed with respect to the matter 
perturbation by the square of the ratio of the perturbation scale $\lambda$ to 
the Hubble radius $r_H= c H^{-1}$ ($H$ being the Hubble constant): 
$\varphi_g/c^2 \sim \delta ~(\lambda / r_H)^2$. 

Such a simplified approach, however, already fails in producing an accurate 
description of the trajectories of relativistic 
particles, such as photons. 
Neglecting the relativistic perturbation of the space--space components 
of the metric, which in the so--called longitudinal gauge is just 
$-2\varphi_g/c^2$, would imply a mistake by a factor of two in 
well--known effects such as the Sachs--Wolfe, Rees--Sciama and 
gravitational lensing, as it would be 
easy to see, by looking at the solution of the eikonal equation. 
In other words, the level of accuracy not only depends on the peculiar 
velocity of the matter producing the spacetime curvature, but also on the 
nature of the particles carrying the signal to the observer. 
Said this way, it may appear that the only relativistic correction required 
to the usual Eulerian Newtonian picture is that of writing the metric tensor 
in the revised, ``weak field", form (e.g. Ref.\cite{bi:peeb93}). 
However, as we are going to show, this is not the whole story. 
It is well--known in fact that 
the gravitational instability of aspherical perturbations (which is the generic 
case) leads to the formation of very anisotropic structures whenever pressure 
gradients can be neglected (e.g. Ref.\cite{bi:shanda}). 
Matter first flows in almost two--dimensional structures called pancakes, 
which then merge and fragment to eventually form one--dimensional filaments 
and point--like clumps. 
During the process of pancake formation the matter density, the shear 
and the tidal field formally become infinite along evanescent 
two--dimensional configurations corresponding to caustics; after this
event a number of highly non--linear phenomena, such as vorticity 
generation by multi--streaming, merging, tidal disruption and 
fragmentation, occur. 
Most of the patology of the caustic formation process, such as the local 
divergence of the density, shear and tide, and the formation of multi--stream 
regions, are just an artifact of extrapolating the pressureless 
fluid approximation beyond 
the point at which pressure gradients and viscosity become important. 
In spite of these limitations, however, it is generally believed that 
the general anisotropy of the collapse configurations, either pancakes or 
filaments, is a generic feature of cosmological structures originated through 
gravitational instability, which would survive even in the presence of a
collisional component. 

This simple observation already shows the inadequacy of the standard Newtonian 
paradigm. According to it, the lowest scale at which the approximation can 
be reasonably applied is set by the amplitude of the gravitational potential 
and is given by the Schwarzschild radius of the collapsing body, which is 
negligibly small for any relevant cosmological mass scale. 
What is completely missing in this criterion is the role of the shear which 
causes the presence of non--scalar contributions to the metric perturbations. 
A non--vanishing shear component is in fact an unavoidable feature of 
realistic cosmological perturbations and affects the dynamics 
in (at least) three ways, all related to non--local effects, i.e. to the 
interaction of a given fluid element with the environment. 

First, at the lowest perturbative order the shear is related to the 
tidal field generated by the surrounding material by a simple proportionality 
law. This sort of non--locality, however,
is coded in the initial conditions of each fluid--element through a 
Coulomb--like interaction with arbitrarily distant matter. Because of 
its link with the initial data of each fluid element one can 
consider it as a {\em local} property. The later modification of these 
shear and tidal fields is one of the consequences of the non--linear evolution. 

Second, it is related to a {\em dynamical} tidal induction: the modification 
of the environment forces the fluid element to modify its shape and density. 
In Newtonian gravity, this is an {\em action--at--a--distance} effect, which 
starts to manifest itself in second--order perturbation theory as an 
inverse--Laplacian contribution to the velocity potential (e.g. Catelan et al. 
\cite{bi:catetal}, and references therein). 

Third, and most important here, a non--vanishing shear field leads to the 
generation of a traceless and divergenceless metric perturbation which can be 
understood as gravitational radiation emitted by non--linear perturbations. 
This contribution to the metric perturbations is statistically 
small on cosmologically interesting scales, but it becomes relevant whenever 
anisotropic (with the only exception of exactly one--dimensional) collapse 
takes place. In the Lagrangian picture considered here, such an effect
already arises at the post--Newtonian (PN) level. 

Note that the two latter effects are only detected if one 
allows for non--scalar perturbations in physical quantities. Contrary to a 
widespread belief, in fact, the choice of scalar perturbations in the initial 
conditions is not enough to prevent tensor modes to arise beyond the linear 
regime in a GR treatment. Truly tensor perturbations are dynamically generated 
by the gravitational instability of initially scalar perturbations, 
independently of the initial presence of gravitational waves. 

Recently a number of different approaches to 
relativistic effects in the non--linear dynamics of cosmological 
perturbations have been proposed. Matarrese, Pantano and Saez 
\cite{bi:mps93} proposed 
an algorithm based on neglecting the magnetic part of the Weyl tensor 
in the dynamics, obtaining strictly local fluid--flow evolution equations, 
i.e. the so--called ``silent universe". Using this formalism 
Bruni, Matarrese and Pantano \cite{bi:silent} 
studied the asymptotic behaviour of 
the system, both for collapse and expansion, showing, in particular, that this 
kind of local dynamics generically leads to spindle singularities for
collapsing fluid elements. This formalism, however, cannot be applied to 
cosmological structure formation {\em inside} the horizon, 
where the non--local tidal 
induction cannot be neglected, i.e. the magnetic Weyl tensor 
$H^\alpha_{~\beta}$ is non--zero, with the exception of highly specific initial 
configurations \cite{bi:mpsprl},\cite{bi:bj94}. 
Rather, it is probably related to the non--linear dynamics of an irrotational 
fluid {\em outside} the (local) horizon \cite{bi:mpsprl},\cite{bi:mps94}.
One possible application \cite{bi:nohair} is in fact 
connected to the so--called {\em Cosmic No--hair Theorem}. 
i.e. to the conjecture that expanding patches of an initially inhomogeneous 
and anisotropic universe asymptotically tend to almost FRW solutions, thanks to
the action of a cosmological constant--like term. 
The self--consistency of these ``silent universe" models has been recently 
demonstrated by Lesame, Dunsby and Ellis \cite{bi:lesame}, 
extending an earlier analysis by Barnes and Rowlingson \cite{bi:barnes}. 
Lesame, Ellis and Dunsby 
\cite{bi:lesamepre} showed that any non--zero $H^\alpha_{~\beta}$ 
has non--vanishing 
divergence (although this appears only as a third--order effect in 
the amplitude of perturbations around the FRW background), reflecting 
the fact that the shear and the electric tide generally have a different 
eigenframe. 
A local--tide approximation for the non--linear evolution of collisionless 
matter, which tries to overcome some limitations of the 
Zel'dovich approximation \cite{bi:zel}, has been recently proposed by 
Hui and Bertschinger \cite{bi:hb95}.  

Following Matarrese and Terranova \cite{bi:pn} we review here a more 
``conservative" approach based on expanding the Einstein and continuity 
equations in inverse powers 
of the speed of light, which will then define a Newtonian limit and, at the 
next order, post--Newtonian corrections. What is new in this approach 
is the choice of the synchronous and comoving gauge, because of which the 
method can be called Lagrangian. 
Various approaches have been proposed in the literature, which are somehow 
related to the present one. A PN approximation has been followed by Futamase 
\cite{bi:futa91}, to describe the dynamics of a clumpy universe. 
Tomita \cite{bi:tomita91} also used non--comoving coordinates in
a PN approach to cosmological perturbations. 
Shibata and Asada \cite{bi:shiba} recently developed a PN approach to 
cosmological perturbations, but they also used non--comoving coordinates. 
Kasai \cite{bi:kasai95} analyzed the non--linear 
dynamics of dust in the synchronous and comoving gauge; his approximation 
methods are however largely different. 
Finally, in a series of papers, based on the Hamilton--Jacobi approach
(e.g. Ref.\cite{bi:salopek} and references therein) 
a new approximation technique has been developed, which relies on an expansion 
in higher and higher gradients of an initial perturbation ``seed".

\section{Relativistic dynamics of collisionless matter} 

In this section we will derive the equations governing the evolution of an
irrotational fluid of dust (i.e. $p=\omega=0$) in a synchronous and comoving 
system of coordinates (actually the possibility of making these two gauge 
choices simultaneously is a peculiarity of irrotational dust, which holds at 
any time,  i.e. also beyond the linear regime). 
The starting point will be the Einstein equations 
$R_{ab} - \frac{1}{2} g_{ab} R = \frac{8\pi G}{c^4} T_{ab}$, with 
$R_{ab}$ the Ricci tensor, and the continuity 
equation $T^{ab}_{~~;a}=0$ for the matter stress--energy tensor 
$T^{ab}=\varrho c^2 u^a u^b$, where $\varrho$ is the mass density and
$u^a$ the fluid four--velocity (normalized to $u^a u_a =-c^2$). 
The line--element reads 
\be
ds^2 = - c^2 dt^2 + h_{\alpha\beta}({\bf q}, t) 
dq^\alpha d q^\beta \;. 
\ee
The fluid four--velocity in comoving coordinates is 
$u^a = (c,0,0,0)$. A fundamental quantity of this analysis will be the 
velocity--gradient tensor, which is purely spatial, 
\be
\Theta^\alpha_{~\beta} \equiv u^\alpha_{~;\beta} = {1 \over 2} 
h^{\alpha\gamma} \dot h_{\gamma\beta} \;, 
\ee
where a dot denotes partial differentiation with respect to the proper time 
$t$. The tensor $\Theta^\alpha_{~\beta}$ represents the {\em extrinsic 
curvature} of the spatial hypersurfaces orthogonal to $u^a$. 

Thanks to the spacetime splitting obtained in our frame, the 10 Einstein 
equations can be immediately divided into 4 constraints and 6 evolution 
equations. The time--time component of the Einstein equations is the 
so--called {\em energy constraint}, which reads 
\be
\Theta^2 - \Theta^\alpha_{~\beta} \Theta^\beta_{~\alpha} + c^2 ~^{(3)}\!R = 
16 \pi G \varrho \;,
\ee
where the {\em volume--expansion scalar} $\Theta$ is just the trace of the 
velocity--gradient tensor, $^{(3)}\!R$ is the trace of the 
three-dimensional Ricci curvature, $^{(3)}\!R^\alpha_{~\beta}$, of the spatial 
hypersurfaces of constant time. 

The space--time components give the {\em momentum constraint}, 
\be 
\Theta^\alpha_{~\beta||\alpha} = \Theta_{,\beta} \;, 
\ee 
where greek indices after a comma denote partial derivatives, while after a 
double vertical bar they denote covariant derivatives in the three--space with 
metric $h_{\alpha\beta}$. 

Finally, the space--space components represent the only truly 
{\em evolution equations}, i.e. those which contain second--order time 
derivatives of the metric tensor. They indeed govern the evolution 
of the extrinsic curvature tensor and read 
\be 
\dot \Theta^\alpha_{~\beta} + \Theta \Theta^\alpha_{~\beta} + 
c^2 ~^{(3)}\!R^\alpha_{~\beta} = 4 \pi G \varrho \delta^\alpha_{~\beta} \;. 
\ee
Taking the trace of the last equation and combining it with the 
energy constraint, one obtains the {\em Raychaudhuri equation},
\be
\dot \Theta + \Theta^\alpha_{~\beta} \Theta^\beta_{~\alpha} + 4 \pi G \varrho
= 0 \;. 
\ee
Mass conservation is provided by the equation
\be 
\dot \varrho = - \Theta \varrho \;. 
\ee

Given that $\Theta = {1 \over 2} h^{\alpha\gamma} \dot h_{\gamma\alpha} =
\partial (\ln h^{1/2}) / \partial t$, where $h \equiv {\rm det} 
~h_{\alpha\beta}$, One can write the solution of this equation in the form 
\be
\varrho({\bf q}, t) = \varrho_0({\bf q}) \bigl[h({\bf q}, t)/ 
h_0 ({\bf q}) \bigr]^{-1/2} \;. 
\ee
Here and in what follows quantities with a subscript $0$ are meant to be 
evaluated at some initial time $t_0$. 

Let us also introduce the so--called {\em electric} and {\em magnetic}
parts of the Weyl tensor, which are both symmetric, flow--orthogonal and 
traceless. They read, respectively,
\be
E^\alpha_{~\beta} = {1 \over 3} \delta^\alpha_{~\beta} \biggl( 
\Theta^\mu_{~\nu} \Theta^\nu_{~\mu} - \Theta^2 \biggr) + \Theta 
\Theta^\alpha_{~\beta} - \Theta^\alpha_{~\gamma} \Theta^\gamma_{~\beta} 
+ c^2 \biggl( ~^{(3)}\!R^\alpha_{~\beta} - {1 \over 3} \delta^\alpha_{~\beta} 
~^{(3)}\!R \biggr)
\ee
and 
\be 
H^\alpha_{~\beta} = {1 \over 2} h_{\beta\mu} \biggl( \eta^{\mu\gamma\delta}
\Theta^\alpha_{~\gamma||\delta} + \eta^{\alpha\gamma\delta}
\Theta^\mu_{~\gamma||\delta} \biggr) \;,
\ee
where $\eta^{\alpha\beta\gamma} h^{-1/2} \epsilon^{\alpha\beta\gamma}$ is the 
three--dimensional, completely anti--symmetric, Levi--Civita tensor 
relative to the spatial metric $h_{\alpha\beta}$ and 
$\epsilon^{\alpha\beta\gamma}$ is such that $\epsilon^{123}=1$, etc... .

Notice that, while the definition of the electric tide $E^\alpha_{~\beta}$ 
is completely fixed, because of its well--known Newtonian limit, 
the magnetic tensor field has no straightforward Newtonian counterpart, and 
can be therefore defined up to arbitrary powers of the speed of light. 
The definition adopted here is the most straightforward one; it 
is such that no explicit powers of $c$ appear, which means that its 
physical dimensions are $1/c$ those of $E^\alpha_{~\beta}$. This choice 
can be motivated in analogy with electrodynamics, where the 
magnetic vector field is also scaled by $1/c$ with respect to the electric 
one. \\

With the purpose of studying gravitational instability in a FRW background, 
it is convenient to factor out the homogeneous and isotropic solutions of 
the above equations, perform a conformal rescaling
of the metric with conformal factor $a(t)$, the scale--factor of
FRW models, and adopt the conformal time $\tau$, 
defined by $d\tau = dt / a(t)$. 

The line--element is then written in the form 
\be
ds^2 = a^2(\tau)\big[ - c^2 d\tau^2 + \gamma_{\alpha\beta}({\bf q}, \tau) 
dq^\alpha d q^\beta \big] \;, 
\ee
where $a^2(\tau) \gamma_{\alpha\beta}({\bf q}, \tau) \equiv 
h_{\alpha\beta}({\bf q}, t(\tau))$. 
For later convenience let us fix the Lagrangian coordinates $q^\alpha$ to have 
physical dimension of length and the conformal time variable $\tau$ to have 
dimension of time. As a consequence the spatial metric $\gamma_{\alpha\beta}$
is dimensionless, as is the scale--factor $a(\tau)$ which must be 
determined by solving the Friedmann equations for a perfect fluid of dust 
\be
\biggl({a' \over a}\biggr)^2 = {8\pi G \over 3} \varrho_b a^2 - \kappa c^2 \;,
\ee
\be 
2 {a'' \over a} - \biggl({a' \over a}\biggr)^2 + \kappa c^2 = 0 \;. 
\ee
Here primes denote differentiation with respect to the conformal time $\tau$
and $\kappa$ represents the curvature parameter of FRW models, 
which, because of our choice of dimensions, cannot be normalized as usual. 
So, for an Einstein--de Sitter universe $\kappa=0$, but for a closed
(open) model one simply has $\kappa>0$ ($\kappa<0$). 
Let us also note that the curvature parameter is related to a 
Newtonian squared time--scale $\kappa_N$ through $\kappa_N \equiv \kappa c^2$
(e.g. Refs.\cite{bi:peeb80},\cite{bi:colluc});in other words $\kappa$ is an 
intrinsically PN quantity. 

By subtracting the isotropic Hubble--flow, one introduces a {\em peculiar 
velocity--gradient tensor}
\be
\vartheta^\alpha_{~\beta} \equiv a \tilde u^\alpha_{~;\beta} - {a' \over a} 
\delta^\alpha_{~\beta} = {1 \over 2} \gamma^{\alpha\gamma} 
{\gamma_{\gamma\beta}}' \;,
\ee
where $\tilde u^a = (c/a,1,1,1)$. 

Thanks to the introduction of this tensor one can rewrite the Einstein's 
equations in a more cosmologically convenient form. 
The energy constraint becomes 
\be
\vartheta^2 - \vartheta^\mu_{~\nu} \vartheta^\nu_{~\mu} + 4 {a' \over a} 
\vartheta + c^2 \bigl( {\cal R} - 6 \kappa \bigr) = 16 \pi G a^2 
\varrho_b \delta \;,
\ee
where ${\cal R}^\alpha_{~\beta}(\gamma) = a^{-2}~^{(3)}\! R^\alpha_{~\beta}(h)$ 
is the conformal Ricci curvature of the three--space, i.e. that corresponding 
to the metric $\gamma_{\alpha\beta}$; for the background FRW solution 
$\gamma^{FRW}_{\alpha\beta} = (1 + {\kappa\over 4} q^2)^{-2} 
\delta_{\alpha\beta}$, one has ${\cal R}^\alpha_{~\beta}(\gamma^{FRW}) 
= 2 \kappa 
\delta^\alpha_{~\beta}$. We also introduced the density contrast 
$\delta \equiv (\varrho - \varrho_b) /\varrho_b$. 

The momentum constraint reads 
\be
\vartheta^\alpha_{~\beta||\alpha} = \vartheta_{,\beta} \;. 
\ee
To avoid excessive proliferation of symbols, the double vertical bars are used 
here and in the following for covariant derivatives in the 
three--space with metric $\gamma_{\alpha\beta}$. 

Finally, after replacing the density from the energy constraint and
subtracting the background contribution, the extrinsic curvature evolution 
equation becomes 
\be
{\vartheta^\alpha_{~\beta}}' + 2 {a' \over a} \vartheta^\alpha_{~\beta} + 
\vartheta \vartheta^\alpha_{~\beta} + {1 \over 4} 
\biggl( \vartheta^\mu_{~\nu} \vartheta^\nu_{~\mu} - \vartheta^2 \biggr) 
\delta^\alpha_{~\beta} + {c^2 \over 4} \biggl[ 4 {\cal R}^\alpha_{~\beta} 
- \bigl( {\cal R} + 2 \kappa \bigr) \delta^\alpha_{~\beta} \biggr]
= 0 \;. 
\ee

The Raychaudhuri equation for the evolution of the 
{\em peculiar volume--expansion scalar} $\vartheta$ becomes 
\be
\vartheta' + {a' \over a} \vartheta + \vartheta^\mu_{~\nu} \vartheta^\nu_{~\mu} 
+ 4 \pi G a^2 \varrho_b \delta =0 \;. 
\ee
The main advantage of this formalism is that there is only one dimensionless 
(tensor) variable in the equations, namely the spatial metric tensor 
$\gamma_{\alpha\beta}$, which is present with its partial time derivatives 
through $\vartheta^\alpha_{~\beta}$, 
and with its spatial gradients through the spatial Ricci 
curvature ${\cal R}^\alpha_{~\beta}$. The only remaining variable is the 
density contrast which can be written in the form
\be
\delta({\bf q}, \tau) = (1 + \delta_0({\bf q})) \bigl[\gamma({\bf q}, \tau)/ 
\gamma_0 ({\bf q}) \bigr]^{-1/2} - 1 \;,
\ee
where $\gamma \equiv {\rm det} ~\gamma_{\alpha\beta}$. 
A relevant advantage of having a single tensorial variable, for our purposes, 
is that there can be no extra powers of $c$ hidden in the definition of 
different quantities. \\

Our intuitive notion of Eulerian coordinates, involving a universal absolute 
time and globally flat spatial coordinates is intimately Newtonian; 
nevertheless it is possible to construct a local coordinates system 
which reproduces this picture for a suitable set of observers. 
This issue has been already addressed in Refs.
\cite{bi:mpsprl},\cite{bi:mps94}, where 
local Eulerian -- FRW comoving -- coordinates $x^A$ where introduced, 
related to the Lagrangian ones $q^\alpha$ via the Jacobian matrix 
with elements
\be
{\cal J}^A_{~~\alpha} ({\bf q},\tau) \equiv 
{\partial x^A \over \partial q^\alpha} \equiv 
\delta^A_{~\alpha} + {\cal D}^A_{~\alpha} ({\bf q},\tau) \;, 
\ \ \ \ \ \ A=1,2,3 \;,
\ee
where ${\cal D}^A_{~\alpha} ({\bf q},\tau)$ is called
{\em deformation tensor}. Each matrix element ${\cal J}^A_{~~\alpha}$
labelled by the Eulerian index $A$ can 
be thought as a three--vector, namely a {\em triad}, defined on the 
hypersurfaces of constant conformal time. 
As shown in Refs.\cite{bi:mpsprl},\cite{bi:mps94}, they evolve according to 
\be
{{\cal J}^A_{~~\alpha}}' = \vartheta^\gamma_{~\alpha} {\cal J}^A_{~~\gamma} \;,
\ee
which also follows from the condition of parallel transport of the triads 
relative to ${\bf q}$ along the world--line of the corresponding 
fluid element $D(a {{\cal J}^A_{~~\alpha}}) / D t =0$. 

Our local Eulerian coordinates are such that the 
spatial metric takes the Euclidean form $\delta_{AB}$,
i.e. 
\be
\gamma_{\alpha\beta} ({\bf q},\tau) = \delta_{AB} {\cal J}^A_{~~\alpha} 
({\bf q},\tau) {\cal J}^B_{~~\beta} ({\bf q},\tau) \;. 
\ee
Correspondingly the matter density can be rewritten in the suggestive form
\be
\varrho ({\bf q},\tau) = \varrho_b(\tau) \bigl( 1 + \delta_0({\bf q}) \bigr)
\bigl[ {\cal J}({\bf q},\tau) / {\cal J}_0 ({\bf q}) \bigr]^{-1} \;,
\ee
where ${\cal J} \equiv {\rm det} {\cal J}^A_{~~\alpha}$. Note that, contrary to
the Newtonian case, it is generally impossible in GR to fix ${\cal J}_0=1$,
as this would imply that the initial Lagrangian space is conformally flat, 
which is only possible if the initial perturbations vanish. \\

We are now ready to deal with the linearization of the equations 
obtained above. This will be done mostly for pedagogical purposes, in 
that it will allow us to obtain a number of results which will turn out 
to be useful for the $1/c^2$ expansion. Apart from this, it can be interesting 
to re--obtain the classical results of linear theory in the comoving and 
synchronous gauge only in terms of the spatial metric coefficients. 

Let us then write the spatial metric tensor of the physical 
(i.e. perturbed) space--time in the form 
\be
\gamma_{\alpha\beta} = {\bar \gamma}_{\alpha\beta} + w_{\alpha\beta} \;,
\ee
with ${\bar \gamma}_{\alpha\beta}$ the spatial metric of the 
background space -- in our case the maximally symmetric FRW one,
${\bar \gamma}_{\alpha\beta} = \gamma^{FRW}_{\alpha\beta}$ -- and 
$w_{\alpha\beta}$ a small perturbation. The only
non--geometric quantity in our equations, namely the initial density contrast
$\delta_0$, can be assumed to be much smaller than unity. 

As usual, one can take advantage of the maximal symmetry of the background FRW 
spatial sections to classify metric perturbations as scalars, vectors and 
tensors. One then writes
\be
w_{\alpha\beta} = \chi {\bar \gamma}_{\alpha\beta} + \zeta_{|\alpha\beta} +
{1 \over 2} \bigl( \xi_{\alpha | \beta} + \xi_{\beta | \alpha} \bigr) 
+ \pi_{\alpha\beta} \;, 
\ee
with
\be
\xi^\alpha_{~~|\alpha}= \pi^\alpha_{~\alpha} = \pi^\alpha_{~\beta|\alpha}=0 \;,
\ee
where a single vertical bar is used for covariant differentiation in the 
background three--space with metric ${\bar \gamma}_{\alpha\beta}$. 
In the above decomposition $\chi$ and $\zeta$ represent scalar modes, 
$\xi^\alpha$ vector modes and $\pi^\alpha_{~\beta}$ tensor modes 
(indices being raised by the contravariant background three--metric). 

Before entering into the discussion of the equations for these perturbation 
modes, let us quote a result which will be also useful later. 
In the $\vartheta^\alpha_{~\beta}$ evolution equation and in the energy 
constraint the combination ${\cal P}^\alpha_{~\beta} \equiv 
4 {\cal R}^\alpha_{~\beta} - \bigl( {\cal R} + 
2 \kappa \bigr) \delta^\alpha_{~\beta}$ and its trace appear. 
To first order in the metric perturbation one has
\be
{\cal P}^\alpha_{~\beta} (w) = -2 \biggl[ \bigl(\nabla^2 - 2 \kappa \bigr) 
\pi^\alpha_{~\beta} + {\chi_|}^\alpha_{~\beta} + \kappa \chi 
\delta^\alpha_{~\beta} \biggr] \;,
\ee
where $\nabla^2 ( \cdot ) \equiv {( \cdot )_|}^\gamma_{~\gamma}$. 
Only the scalar mode $\chi$ and the tensor modes contribute to the 
three--dimensional Ricci curvature. 

As well known, in linear theory scalar, vector and tensor modes are 
independent. The equation of motion for the tensor modes 
is obtained by linearizing the traceless part of the 
$\vartheta^\alpha_{~\beta}$ evolution equation. One has
\be 
{\pi_{\alpha\beta}}'' + 2 {a' \over a} {\pi_{\alpha\beta}}' 
- c^2 \bigl(\nabla^2 - 2 \kappa \bigr) \pi_{\alpha\beta} = 0 \;,
\ee
which is the equation for the free propagation of gravitational 
waves in a FRW background in the Einstein--de Sitter 
case). The general solution of this equation is well--known 
(e.g. Ref.\cite{bi:weinberg}) and will not be reported here. 

At the linear level, in the irrotational case, the two vector modes represent 
gauge modes which can be set to zero, $\xi^\alpha=0$. 

The two scalar modes are linked together through the momentum constraint, 
which leads to the relation
\be
\chi = \chi_0 + \kappa (\zeta - \zeta_0) \;.
\ee
The energy constraint gives
\be
\bigl( \nabla^2 + 3 \kappa \bigr) \biggl[ {a' \over a} \zeta' + 
\bigl(4 \pi G a^2 \varrho_b - \kappa c^2 \bigr) 
(\zeta - \zeta_0) - c^2 \chi_0 \biggr] = 8 \pi G a^2 \varrho_b \delta_0 \;, 
\ee
while the evolution equation gives
\be
\zeta'' + 2 {a' \over a} \zeta' = c^2 \chi \;.
\ee

An evolution equation only for the scalar mode $\zeta$ can be obtained 
by combining together the evolution equation and the energy constraint;
it reads 
\be
\bigl( \nabla^2 + 3 \kappa \bigr) \biggl[ \zeta'' + {a' \over a} \zeta' 
- 4\pi G a^2 \varrho_b (\zeta - \zeta_0) \biggr] = - 8 \pi G a^2 
\varrho_b \delta_0 \;. 
\ee

On the other hand, linearizing the solution of the continuity 
equation, gives 
\be
\delta = \delta_0 - {1 \over 2} (\nabla^2 + 3 \kappa ) (\zeta - \zeta_0) \;,
\ee
which replaced in the previous equation gives 
\be
\delta'' + {a' \over a} \delta' - 4\pi G a^2 \varrho_b \delta = 0 \;.
\ee
This is the well--known equation for linear density fluctuation, whose 
general solution can be found in Ref.\cite{bi:peeb80}. 
Once $\delta(\tau)$ is known, 
one can easily obtain $\zeta$ and $\chi$, which completely solves the linear 
problem. 

Eq.(32) above has been obtained in whole generality; one could have used 
instead the well--known residual gauge ambiguity of the synchronous 
coordinates to simplify its form. In fact, $\zeta$ is determined up to a 
space--dependent scalar, which would neither contribute to the spatial 
curvature, nor to the velocity--gradient tensor. For instance, one
could fix $\zeta_0$ so that $(\nabla^2 + 3 \kappa) \zeta_0 = - 2 \delta_0$,
so that the $\zeta$ evolution equation takes the same form as that for 
$\delta$. 

In order to better understand the physical meaning of the two 
scalar modes $\chi$ and $\zeta$, let us consider the simplest case of an 
Einstein--de Sitter background ($\kappa=0$), for which 
$a(\tau) \propto \tau^2$. By fixing the gauge so that 
$\nabla^2 \zeta_0 = - 2 \delta_0$ one obtains $\chi(\tau)=\chi_0$ and 
\be 
\zeta(\tau) = {c^2 \over 10} \chi_0 \tau^2 + B_0 \tau^{-3} \;,
\ee
where the amplitude $B_0$ of the decaying mode is an arbitrary function of the 
spatial coordinates. Consistency with the Newtonian limit suggests 
$\chi_0 \equiv - {10 \over 3c^2} \varphi_0$, with $\varphi_0$ the initial 
peculiar gravitational potential, related to $\delta_0$ through 
$\nabla^2 \varphi_0 = 4 \pi G a_0^2 \varrho_{0b} \delta_0$. One can then write 
\be
\zeta(\tau) = - {1 \over 3} \varphi_0 \tau^2 + B_0 \tau^{-3} \;.
\ee

This result clearly shows that, at the Newtonian level, the linearized 
metric is $\gamma_{\alpha\beta} = \delta_{\alpha\beta} + \zeta_{|\alpha\beta}$,
while the perturbation mode $\chi$ is already 
PN. Note that also the tensor modes are at least PN. 

These results also confirm the above conclusion that in the general GR case 
the initial Lagrangian spatial metric cannot be flat, i.e. 
${\cal J}_0 \neq 1$, because of the initial ``seed" PN metric perturbation 
$\chi_0$.

\section{Newtonian approximation } 

The Newtonian equations in Lagrangian form can be obtained from the 
full GR equations by an expansion in inverse 
powers of the speed of light; as a consequence of our
gauge choice, however, no odd powers of $c$ appear in the 
equations, which implies that the expansion parameter can be 
taken to be $1/c^2$. 

Let us then expand the spatial metric in a form analogous to that 
used in the linear perturbation analysis of Section 2: 
\be
\gamma_{\alpha\beta} = {\bar \gamma}_{\alpha\beta} + {1 \over c^2} 
w^{(PN)}_{\alpha\beta} + {\cal O} \biggl( {1 \over c^4} \biggr) \;,
\ee
where the $c$ dependence of the metric perturbation was made explicit. 
The actual convergence of the series requires that the 
PN metric perturbation $\frac{1}{c^2} w^{(PN)}_{\alpha\beta}$ is much 
smaller than the background Newtonian metric 
${\bar \gamma}_{\alpha\beta}$. 
Let us first concentrate on the Newtonian metric; the properties of 
$w_{\alpha\beta}$ will be instead considered in Section 4. 

To lowest order in our expansion, the extrinsic curvature evolution equation
and the energy constraint imply that 
${\bar {\cal P}}^\alpha_{~\beta} \equiv 
{\cal P}^\alpha_{~\beta} ({\bar \gamma}) =0$, and recalling that 
$\kappa=\kappa_N/c^2$, 
\be 
{\bar {\cal R}}^\alpha_{~\beta} \equiv {\cal R}^\alpha_{~\beta} ({\bar \gamma}) 
= 0 \;:
\ee 
{\em in the Newtonian limit the spatial curvature identically vanishes}
(e.g. Ref.\cite{bi:ellisvar}). This important conclusion implies that 
${\bar \gamma}_{\alpha\beta}$ can be transformed to $\delta_{AB}$
globally, i.e. that one can write 
\be
{\bar \gamma}_{\alpha\beta} = \delta_{AB} {\bar {\cal J}}^A_{~~\alpha} 
{\bar {\cal J}}^B_{~~\beta} \;,
\ee
with integrable Jacobian matrix coefficients. In other words, at each time 
$\tau$ there exist {\em global Eulerian coordinates} $x^A$ such that 
\be
{\bf x}({\bf q}, \tau) = {\bf q} + {\bf S}({\bf q}, \tau) \;,
\ee
where ${\bf S}({\bf q}, \tau)$ is called the {\em displacement vector}, 
and the deformation tensor becomes in this limit 
\be
{\bar {\cal D}}^A_{~\alpha} = {\partial S^A \over \partial q^\alpha} \;.
\ee

The Newtonian Lagrangian metric can therefore be written in the form 
\be
{\bar \gamma}_{\alpha\beta}({\bf q}, \tau) = \delta_{AB} 
\biggl(\delta^A_{~\alpha} + {\partial S^A({\bf q}, \tau) 
 \over \partial q^\alpha} \biggr) 
\biggl(\delta^B_{~\beta} + {\partial S^B({\bf q}, \tau) 
\over \partial q^\beta} \biggr) \;.
\ee 

One can rephrase the above result as follows: the Lagrangian spatial metric in 
the Newtonian limit is that of Euclidean three--space in time--dependent 
curvilinear coordinates $q^\alpha$, defined at each time $\tau$ in terms of 
the Eulerian ones $x^A$ by inverting Eq.(40) above. As a consequence, the 
Christoffel symbols involved in spatial covariant derivatives (which will be
indicated by a single bar or by a nabla operator followed by greek indices) 
do not vanish, but the vanishing of the spatial curvature implies that 
these covariant derivatives always commute. 

Contrary to the evolution equation and the energy constraint, the 
Raychaudhuri equation and the momentum constraint 
contain no explicit powers of $c$, and therefore preserve their form in going 
to the Newtonian limit. These equations therefore determine the background 
Newtonian metric ${\bar \gamma}_{\alpha\beta}$, i.e. they govern the evolution 
of the displacement vector ${\bf S}$. 

The Raychaudhuri equation becomes the master equation for the Newtonian 
evolution; it takes the form 
\be 
{\bar \vartheta}' + {a' \over a} {\bar \vartheta} + 
{\bar \vartheta}^\mu_{~\nu} {\bar \vartheta}^\nu_{~\mu} + 
4 \pi G a^2 \varrho_b \bigl( {\bar \gamma}^{-1/2} - 1 \bigr) = 0 \;,
\ee
where
\be
{\bar \vartheta}^\alpha_{~\beta} \equiv {1 \over 2} 
{\bar \gamma}^{\alpha\gamma} {\bar \gamma_{\gamma\beta}}' \;, 
\ee
and, for simplicity, $\delta_0=0$ was assumed (a restriction which is, however,
not at all mandatory). We also used the residual gauge freedom 
of our coordinate system to set ${\bar \gamma}_{\alpha\beta}(\tau_0) = 
\delta_{\alpha\beta}$, implying ${\bar {\cal J}}_0=1$, i.e. to make 
Lagrangian and Eulerian coordinates coincide at the initial time. 
That this choice is indeed possible in the Newtonian limit can be understood 
from our previous linear analysis, where this is achieved by taking, e.g., 
$\zeta_0=0$. 

The momentum constraint, 
\be
{\bar \vartheta}^\mu_{~\nu |\mu } = {\bar \vartheta}_{,\nu} \;, 
\ee
is actually related to the irrotationality assumption. 

Let us also notice a general property of our 
expression for the Lagrangian metric: at each time $\tau$ it can be 
diagonalized by going to the local and instantaneous principal axes
of the deformation tensor. Calling $\bar \gamma_\alpha$ 
the eigenvalues of the 
metric tensor, ${\bar {\cal J}}_\alpha$ those of the Jacobian and 
${\bar d}_\alpha$ those of the deformation tensor, 
one has
\be
\bar \gamma_\alpha ({\bf q},\tau) = {\bar {\cal J}}_\alpha^2({\bf q},\tau) = 
\bigl( 1 + {\bar d}_\alpha ({\bf q},\tau) \bigr)^2 \;.
\ee

The diagonal form of the metric tensor will be 
reconsidered in the frame of the Zel'dovich approximation. Beyond the 
mildly non--linear regime, where this approximation is consistently applied, 
diagonalizing the metric is in general, i.e. apart from 
specific initial configurations, of smaller practical use, because 
metric (and deformation) tensor, shear and tide generally have different 
eigenvectors. 

From this expression it becomes evident that, at shell--crossing, 
where some of the Jacobian eigenvalues go to zero, the related covariant 
metric eigenvalues just vanish. On the other hand, other quantities, like 
the matter density, the peculiar volume expansion scalar and some eigenvalues 
of the shear and tidal tensor will generally diverge at the location of the 
caustics (see Ref.\cite{bi:silent}, for a discussion). 
This diverging behaviour makes the description of the system 
extremely involved after this event. Although dealing with this problem is 
far outside the aim of the present paper, let us just mention that a number of 
ways out are available. One can convolve the various dynamical variables by a 
suitable low--pass filter, either at the initial time, in order to postpone 
the occurrence of shell--crossing singularities, or at the time when they 
form, in order to smooth the singular behaviour; alternatively one 
can abandon the perfect fluid picture and resort to a discrete point--like 
particle set, which automatically eliminates the possible occurrence of 
caustics, at least for generic initial data. 
At this level, anyway, we prefer to take a conservative 
point of view and assume that the  actual range of validity of our formalism 
is up to shell--crossing. \\

A more direct way to deal with the Lagrangian Newtonian equations is
to write them in terms of the Jacobian matrix ${\cal J}^A_{~~\alpha}$.
This approach is obviously related to the more usual ones in terms of 
the displacement vector ${\bf S}$ or in terms of the deformation tensor 
${\cal D}^A_{~~\alpha}$ \cite{bi:buc89},\cite{bi:moutarde},\cite{bi:bjcp},
\cite{bi:buc92},\cite{bi:lach},\cite{bi:catelan}. 

In order to rewrite the Raychaudhuri equation in terms of the Jacobian matrix, 
we notice that 
\be
{\vartheta^\alpha_{~\beta}}' = {\cal J}^\alpha_{~~A} {\cal J}^A_{~~\beta}~''
- \vartheta^\alpha_{~\mu} \vartheta^\mu_{~\beta} \;,
\ee
where we have introduced the inverse Jacobian matrix
\be
{\cal J}^\alpha_{~~A} \equiv {\partial q^\alpha \over \partial x^A} \;, 
\ee
where Eulerian indices are raised and lowered by the Kronecker symbol. 
To make explicit the notation, we just stress that elements of 
${\partial x^A \over \partial q^\alpha}$ will be characterized 
by a greek (i.e. Lagrangian) index subscript, while elements of the 
inverse matrix ${\partial q^\alpha \over \partial x^A}$ will 
be characterized by a greek index superscript. 

Replacing the latter identity into the Newtonian expression Eq.(43) yields
\be
{\bar {\cal J}}^\alpha_{~A} \bar {{\cal J}^A_{~~\alpha}}'' + 
{a' \over a} {\bar {\cal J}}^{-1} {\bar {\cal J}}' = 
4 \pi G a^2 \varrho_b \bigl( 1 - {\bar {\cal J}}^{-1} \bigr) \;. 
\ee

We also notice that the parallel transport condition can
be rewritten in the form 
\be
\vartheta^\alpha_{~\beta} = {\cal J}^\alpha_{~A} {{\cal J}^A_{~~\beta}}' \;.
\ee

This equation, together with $\vartheta^\alpha_{~\beta} = \frac{1}{2} 
\gamma^{\alpha\gamma} {\gamma_{\gamma\beta}}'$ gives the general relation 
\be
{{\cal J}_{A\alpha}}' {\cal J}^A_{~~\beta} = {\cal J}_{A\alpha} 
{{\cal J}^A_{~~\beta}}' \;.
\ee
Replacing these relations in the momentum constraint we obtain 
in whole generality 
\be
{\cal J}^\alpha_{~~A} {{{\cal J}^A_{~~\beta}}'}_{||\alpha} +
{{\cal J}^\alpha_{~~A}}_{||\alpha} {{\cal J}^A_{~~\beta}}' = 
\bigl({\cal J}^{-1} {\cal J}' \bigr)_{,\beta} \;.
\ee

On the other hand, in the Newtonian limit we have 
\be
{\bar {\cal J}}^A_{~~\alpha,\beta} = {\bar {\cal J}}^A_{~~\beta,\alpha} \;, 
\ee
as it follows from the fact that $S^A_{~~,\alpha\beta} 
= S^A_{~~,\beta\alpha}$. Using this commutation property it is easy to 
verify that 
\be
{\bar \Gamma}^\alpha_{\beta\gamma} = {\bar {\cal J}}^\alpha_{~A} 
{\bar {\cal J}}^A_{~~\beta,\gamma} \;.
\ee 
Thanks to the latter relation and to the well--known matrix identity 
${\rm Tr} \ln {\bf J} = \ln {\rm det} {\bf J}$, it is straightforward to 
verify that the momentum constraint in the Newtonian limit becomes an 
identity. 
It is then clear that Eq.(51) is more fundamental than the momentum 
constraint:
it plays the role of an irrotationality condition written in Lagrangian space. 
This is of course equivalent to the standard form [compare with 
Eq.(59) in Ref.\cite{bi:catelan} $\epsilon^{\alpha\beta\gamma} 
{\bar {\cal J}}^A_{~~\beta} {{\bar {\cal J}}_{A\gamma}}'$.
This equation, together with the Raychaudhuri equation above, 
completely determines the Newtonian problem, in terms of either the Jacobian 
matrix, the deformation tensor or the displacement vector. 

As demonstrated above, it is always 
possible, in the frame of the Newtonian approximation, to define a global 
Eulerian picture. This will be the picture of the fluid evolution as given 
by an observer that, at the point ${\bf x} = {\bf q} + {\bf S}({\bf q}, \tau)$ 
and at the time $\tau$ observes the fluid moving with physical peculiar 
three--velocity ${\bf v} = d {\bf S}/d \tau$. From the point
of view of a Lagrangian observer, who is comoving with the fluid, 
the Eulerian observer, which is located at constant ${\bf x}$, is moving with 
three--velocity $d {\bf q} ({\bf x}, \tau)/d \tau = - {\bf v}$. 

The line--element characterizing the Newtonian approximation in the 
Eulerian frame is well--known (e.g. Ref.\cite{bi:peeb80}) 
\be
ds^2 = a^2(\tau) \biggl[ - \biggl(1 + {2\varphi_g ({\bf x}, \tau) \over c^2} 
\biggr) ~c^2 d\tau^2 + \delta_{AB} dx^A dx^B \biggr] \;, 
\ee
with $\varphi_g$ the peculiar gravitational potential, determined 
by the mass distribution through the Eulerian Poisson equation, 
\be
\nabla_x^2 \varphi_g ({\bf x}, \tau) = 4 
\pi G a^2(\tau) \varrho_b(\tau) \delta({\bf x}, \tau) \;,
\ee
where the Laplacian $\nabla_x^2$, as well as the nabla operator $\nabla$, 
have their standard Euclidean meaning. 
The perturbation in the time--time component of the metric tensor 
here comes from the different proper time of the Eulerian 
and Lagrangian observers. 

It is now crucial to realize that all the dynamical equations obtained 
so far, being entirely expressed in terms of three--tensors, keep their form 
in going to the Eulerian picture, only provided the convective time 
derivatives of tensors of any rank (scalars, vectors and 
tensors) are modified as follows: 
\be
{D \over D \tau} \rightarrow {\partial \over \partial \tau} + 
{\bf v} \cdot \nabla \;, \ \ \ \ \ \ \ 
{\bf v} \equiv {d {\bf S} \over d \tau} \;.
\ee

This follows from the fact that, for the metric above, $\bar \Gamma^0_{AB} = 
\bar \Gamma^A_{0B} =\bar \Gamma^A_{BC}=0$, which also obviously implies that 
covariant derivatives with respect to $x^A$ reduce to partial ones. 

The irrotationality assumption now has the obvious consequence that we can 
define an Eulerian velocity potential $\Phi_v$ through
\be
{\bf v} ({\bf x}, \tau) = \nabla \Phi_v ({\bf x}, \tau) \;.
\ee
The Newtonian peculiar velocity--gradient tensor then becomes 
\be
\bar \vartheta_{AB} = {\partial^2 \Phi_v \over \partial x^A \partial x^B} \;,
\ee
because of which the momentum constraint gets trivially satisfied and the 
magnetic Weyl tensor becomes identically zero in the Newtonian limit. 

We can now write the Raychaudhuri equation for the Eulerian peculiar 
volume--expansion scalar $\bar \vartheta$, 
and use the Poisson equation to get, as a first spatial integral, the 
{\em Euler equation} 
\be
{\bf v}~' + {\bf v} \cdot \nabla {\bf v} + 
{a' \over a} {\bf v} = - \nabla \varphi_g \;.
\ee

This can be further integrated to give the {\em Bernoulli equation} 
\be
\Phi_v' + {a' \over a} \Phi_v + {1\over 2} \bigl(\nabla \Phi_v \bigr)^2 
= - \varphi_g \;.
\ee

On the other hand, by taking gradients of the Euler equation we can 
obtain an Eulerian evolution equation for the tensor $\bar \vartheta_{AB}$.
More interesting is that this equation can be transported back to the 
Lagrangian frame to get 
\be
{\bar \vartheta}^{\alpha~'}_{~\beta}
+ {a' \over a} \bar \vartheta^\alpha_{~\beta}  + 
\bar \vartheta^\alpha_{~\gamma}  
\bar \vartheta^\gamma_{~\beta}  = - {\varphi_{g|}^{(L)}}^\alpha_{~\beta} \;. 
\ee
where $\varphi_g^{(L)}$ must be thought as a {\em Lagrangian peculiar 
gravitational potential} to be 
determined through the {\em Lagrangian Poisson equation} 
\be
{\varphi_{g|}^{(L)}}^\alpha_{~\alpha} = 4 \pi G a^2 \varrho_b 
(\bar \gamma^{-1/2} -1) \;.
\ee

These two Lagrangian expressions will turn out to be very useful for the
PN calculations of Section 4. \\

Having shown the equivalence of this method, in the Newtonian limit, with 
the standard one, it is now trivial to recover the Zel'dovich approximation 
\cite{bi:zel}. This is obtained by expanding Eq.(49) and Eq.(51) to first 
order in the displacement vector. The result is 
\be
{\bf x} ({\bf q},\tau) = {\bf q} + D(\tau) \nabla \Phi_0({\bf q}) \;,
\ee
where only the growing mode solution $D(\tau)$ of Eq.(34) has 
been considered, 
and we introduced the potential $\Phi_0({\bf q})$, such that 
$\nabla_q^2 \Phi_0 = - \delta_0/D_0$, where $\nabla_q^2$ is the standard 
(i.e. Euclidean) Laplacian in Lagrangian coordinates; more in general, 
at this perturbative order covariant and partial derivatives with respect
to the $q^\alpha$ coincide. The potential $\Phi_0$ is easily related 
to the initial peculiar gravitational potential defined in the Introduction, 
$\Phi_0 = - (4\pi G a_0^2 \varrho_{0b} D_0)^{-1} \varphi_0$. 

More interesting is to derive from the above expression the 
corresponding {\em Zel'dovich metric}. It reads 
\be
\gamma^{ZEL}_{\alpha\beta} ({\bf q},\tau) = \delta_{\gamma\delta} 
\biggl(\delta^\gamma_{~\alpha} + D(\tau) {\Phi_0,}^\gamma_{~\alpha} 
({\bf q}) \biggr) 
\biggl(\delta^\delta_{~\beta} + D(\tau) {\Phi_0,}^\delta_{~\beta} 
({\bf q}) \biggr) \;.
\ee

One can of course diagonalize this expression by going to the principal axes 
of the deformation tensor. Calling $\lambda_\alpha$ 
the eigenvalues of the matrix ${\Phi_0,}^\alpha_{~\beta}$, one finds 
\be
\gamma^{ZEL}_\alpha ({\bf q}, \tau) = \bigl[ 1 + D(\tau) 
\lambda_\alpha({\bf q})\bigr]^2  \;.
\ee

Note that, contrary to what has been commonly done so far in the literature, 
the metric tensor must be evaluated at second order in the displacement vector, 
in order to obtain back the correct Zel'dovich expressions for the dynamical 
variables (density, shear, etc ...). 

The above diagonal form of the metric allows a straightforward calculation of 
all the relevant quantities. The well--known expression for the mass density 
is consistently recovered, 
\be
\varrho^{ZEL} = \varrho_b \prod_\alpha
\bigl(1 + D \lambda_\alpha \bigr)^{-1} \;.
\ee
The peculiar velocity--gradient tensor has the same eigenframe of 
the metric; its eigenvalues read 
\be
\vartheta^{ZEL}_\alpha = {D' \lambda_\alpha 
\over 1 + D \lambda_\alpha } \;.
\ee
By summing over $\alpha$ the latter expression we can obtain the 
peculiar volume--expansion scalar 
\be
\vartheta^{ZEL} = \sum_\alpha 
{D' \lambda_\alpha \over 1 + D \lambda_\alpha} 
\ee
and then the shear eigenvalues 
\be 
\sigma^{ZEL}_\alpha = {D' \lambda_\alpha \over 1 + D \lambda_\alpha} -
{1 \over 3} \sum_\alpha 
{D' \lambda_\alpha \over 1 + D \lambda_\alpha} \;.
\ee
The (conformally rescaled) electric tide ${\cal E}^\alpha_{~\beta} \equiv a^2 
E^\alpha_{~\beta}$ comes out just proportional to the shear. 
Its eigenvalues read 
\be
{\cal E}^{ZEL}_\alpha = - 4 \pi G a^2 \varrho_b {D \over D'} 
~\sigma^{ZEL}_\alpha \;. 
\ee

These expressions for the shear and the tide completely agree with those 
obtained by Kofman and Pogosyan \cite{bi:kp95} and Hui and Bertschinger 
\cite{bi:hb95}. 
The fact that metric, shear and tide have simultaneous eigenvectors shows that 
fluid elements in the Zel'dovich approximation actually evolve in a way much 
similar to a ``silent universe" \cite{bi:mps93},\cite{bi:silent}, i.e. 
with no influence from the environment, except for that implicit in the 
self--consistency of the initial conditions. 

\section{Beyond the Newtonian approximation }

Having examined all the aspects of the formalism in the Newtonian 
limit, we are now ready to proceed to the next perturbative order in $1/c^2$. 
The PN terms $\frac {1}{c^2} w^{(PN)}_{\alpha\beta}$ in Eq.(37) should be 
thought as small perturbations superposed on a Newtonian background
$\bar \gamma_{\alpha\beta}$. 
The fact that the three--metric in the Newtonian limit is that of
Euclidean space in time--dependent curvilinear coordinates $q^\alpha$, implies
that we can apply most of the standard tools of linear perturbation theory in
a flat spatial background (actually, in an Einstein--de Sitter universe). In 
particular, we can classify the PN metric perturbations as 
scalar, vector and tensor modes, as usual. 

We then write
\be
w^{(PN)}_{\alpha\beta} = 
\chi^{(PN)} {\bar \gamma}_{\alpha\beta} + \zeta^{(PN)}_{|\alpha\beta} +
{1 \over 2} \bigl( \xi^{(PN)}_{\alpha | \beta} + \xi^{(PN)}_{\beta | \alpha} 
\bigr) + \pi^{(PN)}_{\alpha\beta} \;, 
\ee
with
\be
{\xi^{(PN)}}^\alpha_{~~|\alpha}= {\pi^{(PN)}}^\alpha_{~\alpha} = 
{\pi^{(PN)}}^\alpha_{~\beta|\alpha}=0 \;,
\ee
where greek indices after a single vertical bar, or nabla operators with a 
greek index, denote covariant differentiation in the 
Newtonian background three--space with metric ${\bar \gamma}_{\alpha\beta}$. 
In the above decomposition $\chi^{(PN)}$ and $\zeta^{(PN)}$ represent PN 
scalar modes, $\xi^{(PN)}_\alpha$ PN vector modes and 
$\pi^{(PN)}_{\alpha\beta}$ PN tensor ones 
(indices being raised by the contravariant background three--metric). 
We deliberately used the same symbols as in Section 2, in order to 
emphasize the analogy with the linear problem. 
Some of these PN modes, namely $\chi^{(PN)}$ and 
$\pi^{(PN)}_{\alpha\beta}$, also have a non--vanishing linear counterpart,
as noticed in Section 2 (actually the linear part of 
$\pi^{(PN)}_{\alpha\beta}$ appears as a gauge mode in the equations), 
while others, namely $\zeta^{(PN)}$, and $\xi_\alpha^{(PN)}$ are intrinsically 
non--linear. Unlike linear perturbation theory in a FRW background, 
metric perturbations of different rank do not decouple: this is because 
our time--dependent Newtonian background enters the 
equations not only through the metric $\bar \gamma_{\alpha\beta}$, but also 
through the peculiar velocity--gradient tensor 
$\bar \vartheta^\alpha_{~\beta}$, which also contains scalar, vector and tensor 
modes. This fact leads to non--linear scalar--vector, 
scalar--tensor and vector--tensor mode mixing, which also explains why we
had to account for the vector modes $\xi^{(PN)}_\alpha$ in the expansion of 
$w^{(PN)}_{\alpha\beta}$, in spite of the irrotational character of our fluid 
motions. 

As in every perturbative calculation, some of the equations 
have the property to mix different perturbative orders. This is of course 
necessary in order to make the $n$--th order coefficients of the expansion 
calculable in terms of those of order $n-1$. In our case the energy constraint
and the extrinsic curvature evolution equation (which at the Newtonian level 
implies ${\cal R}^\alpha_{~\beta}(\bar \gamma)=0$) play this role. 
Therefore we assume that the Newtonian metric and its derivatives are known 
by solving the Raychaudhuri equation and the momentum constraint, 
and we calculate the PN metric perturbations in terms of them. 

Let us first compute the tensor ${\cal P}^\alpha_{~\beta} \equiv 
4 {\cal R}^\alpha_{~\beta} - \bigl( {\cal R} + 2 \kappa \bigr) 
\delta^\alpha_{~\beta}$ to first order in $1/c^2$. We obtain 
\be
c^2 {\cal P}^\alpha_{~\beta} (w^{(PN)})  = - 2 \biggl( 
\nabla^2 {\pi^{(PN)}}^\alpha_{~\beta} + {\chi^{(PN)}_|}^\alpha_{~\beta} 
\biggr) \;, 
\ee
where now $\nabla^2 ( \cdot ) \equiv {( \cdot )_|}^\alpha_{~\alpha}$.

In the energy constraint only the scalar $\chi$ enters, 
\be
\nabla^2 \chi = {1 \over 2} \biggl( \bar \vartheta^2 - \bar 
\vartheta^\mu_{~\nu} \bar \vartheta^\nu_{~\mu} \biggr) + 
2 {a' \over a} \bar \vartheta - 8 \pi G a^2 \varrho_b 
\bigl( \bar \gamma^{-1/2} - 1 \bigr) \;, 
\ee
where, here and from now on, we have dropped the superscript (PN) 
on PN terms. 
One can also obtain an equation for $\chi$ from the trace of the evolution 
equation, Eq.(17), which is however equivalent to the latter, 
thanks to the Newtonian Raychaudhuri equation. 

The tensor perturbations $\pi^\alpha_{~\beta}$ are instead determined via
the evolution equation, Eq.(17) (actually from its trace--free part), 
\be
\nabla^2 \pi^\alpha_{~\beta} = 2 \bar {\vartheta^\alpha_{~\beta}}' 
+ 2 \biggl( 2 {a' \over a} + \bar \vartheta \biggr) 
\bar \vartheta^\alpha_{~\beta} - 
{1 \over 2} \delta^\alpha_{~\beta} 
\biggl( \bar \vartheta^2 - 
\bar \vartheta^\mu_{~\nu} \bar \vartheta^\nu_{~\mu} \biggr) - 
{\chi_|}^\alpha_{~\beta} \;. 
\ee

A by--product of the latter equation is that linear tensor modes, which in 
the $c \to \infty$ limit appear as harmonic functions (i.e. pure gauge modes), 
do not contribute to the r.h.s., i.e. to the Newtonian evolution of the 
system, as expected. 

In order to get an equation for the tensor modes decoupled from the scalar 
mode $\chi$ we can resort to the equations obtained at the end of Section 3. 
To this aim we define the auxiliary Newtonian potential $\Psi_v^{(L)}$, through 
\be
\nabla^2 \Psi_v^{(L)} = - {1 \over 2} \biggl( \bar \vartheta^2 - \bar 
\vartheta^\mu_{~\nu} \bar \vartheta^\nu_{~\mu} \biggr) \;.
\ee

Using this definition in Eq.(75), we obtain 
\be
\chi = - \Psi_v^{(L)} + 2 {a' \over a} \Phi_v^{(L)}- 2 \varphi_g^{(L)} \; 
\ee
(which, at the linear level reduces to the expression of Section 2, 
$\chi_0= - \frac{10}{3c^2} \varphi_0$).  
Using this expression in Eq.(76) and replacing 
$\bar {\vartheta^\alpha_{~\beta}}'$ from Eq.(62) we obtain
\be
\nabla^2 \pi^\alpha_{~\beta} = \biggl( \nabla^\alpha \nabla_\beta 
+ \delta^\alpha_{~\beta} \nabla^2 \biggr) \Psi_v^{(L)} + 
2 \biggl( \bar \vartheta \bar \vartheta^\alpha_{~\beta} - 
\bar \vartheta^\alpha_{~\gamma} 
\bar \vartheta^\gamma_{~\beta} \biggr) \;, 
\ee
which has the significant advantage of being explicitly second 
order (in any possible perturbative approach). 
This equation is one of the most important results of this paper: 
it gives (in the so--called {\em near zone}) the amount of gravitational waves 
emitted by non--linear cosmological perturbations, 
evolved within Newtonian gravity. In other terms, this
equation, which is only applicable on scales well inside 
the horizon, describes gravitational waves produced by an 
inhomogeneous Newtonian background. 

We have 
$\bar \vartheta^\alpha_{~\beta} ({\bf q}, \tau)= D'(\tau) 
{\Phi_{0,}}^\alpha_{~\beta}({\bf q})$, 
with $D(\tau)$ the growing mode solution of Eq.(34) ($D(\tau) \propto a(\tau) 
\propto \tau^2$ in the Einstein--de Sitter case) and $\Phi_0$ defined in 
Section 3. Therefore 
\be
\nabla_q^2 \pi^\alpha_{~\beta} = {D'}^2 \biggl[ 
{\Psi_{0,}}^\alpha_{~\beta} + \delta^\alpha_{\beta} 
\nabla_q^2 \Psi_0 + 2 \biggl( {\Phi_{0,}}^\alpha_{~\beta} 
\nabla_q^2 \Phi_0 - {\Phi_{0,}}^\alpha_{~\gamma}  
{\Phi_{0,}}^\gamma_{~\beta} \biggr) \biggr] \;, 
\ee
where the symbol $\nabla^2_q$ indicates the standard (Euclidean) form of the 
Laplacian in Lagrangian coordinates and $\Psi_0 \equiv \Psi_v(\tau_0)$ and 
indices are raised by the Kronecker symbol. 

To completely determine the PN metric perturbations we still need 
the scalar mode $\zeta$ and the vector modes $\xi_\alpha$, which can be
computed through the momentum constraint. 
We then need $\vartheta^\alpha_{~\beta}$ at the PN order; it reads
\be
\vartheta^\alpha_{~\beta} = \bar \vartheta^\alpha_{~\beta} + {1 \over c^2}
\biggl( \bar \vartheta^\alpha_{~\gamma} w^\gamma_{~\beta} - 
\bar \vartheta^\gamma_{~\beta} w^\alpha_{~\gamma} + 
{1 \over 2} {w^\alpha_{~\beta}}' \biggr) \;. 
\ee
By replacing this expression into the momentum constraint, and 
using the Newtonian identity $\bar {\Gamma^\mu_{\nu\rho}}' = 
\bar \vartheta^\mu_{~\nu|\rho}$, one can obtain 
\begin{eqnarray}
\nonumber
2 {\chi_{,\beta}}' + 
\bar \vartheta \chi_{,\beta} - 
3 \bar \vartheta^\alpha_{~\beta} \chi_{,\alpha} - 
\bar \vartheta_{,\alpha} \pi^\alpha_{~\beta} + 
\bar \vartheta^\gamma_{~\beta|\alpha} \pi^\alpha_{~\gamma} - 
2 \bar \vartheta^\alpha_{~\gamma} \pi^\gamma_{~\beta|\alpha} 
+ \bar \vartheta^\alpha_{~\gamma} \pi^\gamma_{~\alpha|\beta} -
\\
\nonumber
 - \bar \vartheta_{,\alpha} {\zeta_|}^\alpha_{~\beta} + 
\bar \vartheta^\gamma_{~\beta|\alpha} {\zeta_|}^\alpha_{~\gamma} -
\bar \vartheta^\alpha_{~\gamma} {\zeta_|}^\gamma_{~\beta\alpha} + 
\bar \vartheta^\alpha_{~\beta} {\zeta_|}^\gamma_{~\alpha\gamma} - 
{1 \over 2} \bigl(\nabla^2 \xi_\beta \bigr)' - 
{1 \over 2} \bar \vartheta_{,\alpha} \xi^\alpha_{~|\beta} -
\\
 - {1 \over 2} \bar \vartheta_{,\alpha} \xi_{\beta|}^{~~\alpha} + 
\bar \vartheta^\gamma_{~\beta|\alpha} \xi^\alpha_{~|\gamma} - 
\bar \vartheta^\alpha_{~\gamma} {\xi_{\beta|}}^\gamma_{~\alpha} +
\bar \vartheta^\alpha_{~\beta} \nabla^2 \xi_\alpha = 0 \;, 
\end{eqnarray}
which shows that $\zeta$ and $\xi^\alpha$ are 
implicitly determined by the Newtonian quantities, once $\chi$ and 
$\pi^\alpha_{~\beta}$ have been computed. 

\section{Fluid--flow approach }

Following the {\em fluid--flow} approach, described in the classical review by 
Ellis \cite{bi:ellisvar} (see also Ref.\cite{bi:ehlers}), we 
can alternatively describe our system 
in terms of fluid properties, in our case matter density, 
volume--expansion scalar and shear tensor, and two geometric tensors, the 
electric and magnetic parts of the Weyl tensor defined above. The 
derivation of the equations reported below is thoroughly described by 
Ellis \cite{bi:ellisvar} and will not be reported here. 

For most cosmological purposes it is convenient to adopt the conformal 
rescaling and FRW background subtraction described in Section 2. 
Therefore, we can start by writing the continuity equation directly in terms 
of the density contrast $\delta$,
\be 
{D \delta \over D \tau} + \bigl(1 + \delta \bigr) \vartheta = 0 \;, 
\ee
with $D \over D \tau$ denoting convective differentiation with respect to the 
conformal time $\tau$. In the Lagrangian frame, however, and for a scalar 
field, convective differentiation and partial differentiation coincide. 
The peculiar volume--expansion scalar $\vartheta$ obeys 
the Raychaudhuri equation 
which we can rewrite in the form 
\be 
{D \vartheta \over D \tau} + {a' \over a} \vartheta + {1 \over 3} \vartheta^2
+ \sigma^\alpha_{~\beta} \sigma^\beta_{~\alpha} + 
4 \pi G a^2 \varrho_b \delta = 0 \;,
\ee
where $\sigma^\alpha_{~\beta} \equiv \vartheta^\alpha_{~\beta} - 
{1 \over 3} \delta^\alpha_{~\beta} \vartheta$ 
is the shear tensor. The shear, in turn, evolves according to 
\be
{D \sigma^\alpha_{~\beta} \over D \tau} + {a' \over a} \sigma^\alpha_{~\beta} 
+ {2 \over 3} \vartheta \sigma^\alpha_{~\beta} + \sigma^\alpha_{~\gamma} 
\sigma^\gamma_{~\beta} - {1 \over 3} \delta^\alpha_{~\beta} 
\sigma^\gamma_{~\delta} \sigma^\delta_{~\gamma} + {\cal E}^\alpha_{~\beta} = 
0 \;,
\ee
where we have rescaled the electric tide as ${\cal E}^\alpha_{~\beta} \equiv 
a^2 E^\alpha_{~\beta}$, which can be written in terms of our
new variables as 
\be
{\cal E}^\alpha_{~\beta} = {1 \over 3} \delta^\alpha_{~\beta} 
\sigma^\mu_{~\nu} \sigma^\nu_{~\mu} + {1 \over 3} \vartheta 
\sigma^\alpha_{~\beta} + {a' \over a} \sigma^\alpha_{~\beta} - 
\sigma^\alpha_{~\gamma} \sigma^\gamma_{~\beta} 
+ c^2 \biggl( {\cal R}^\alpha_{~\beta} - {1 \over 3} \delta^\alpha_{~\beta} 
{\cal R} \biggr) \;.
\ee

Note that, for a generic second rank tensor $A^\alpha_{~\beta}$, one has 
\be
{D A^\alpha_{~\beta} \over D \tau} = {d A^\alpha_{~\beta} \over d \tau} + 
\sigma^\alpha_{~\gamma} A^\gamma_{~\beta} - \sigma^\gamma_{~\beta} 
A^\alpha_{~\gamma} \;,
\ee
where ${d \over d \tau}$ denotes the total derivative with respect to $\tau$,
which in comoving coordinates coincides with the partial one. 
The two last terms in the r.h.s. come from writing the 
Christoffel symbols in our gauge. It is then clear 
that when the ${D \over D \tau}$ operator acts on either the shear or 
the complete $\vartheta^\alpha_{~\beta}$ tensor, the second and third term 
in the r.h.s. cancel each other and the convective and total 
differentiation coincide. 
This cancellation also occurs
for a generic $A^\alpha_{~\beta}$ if either the relevant Christoffel symbols 
vanish (as it is the case for the Newtonian limit in Eulerian coordinates) or 
the convective derivative acts on the 
eigenvalues of $A^\alpha_{~\beta}$ and such a tensor has the same eigenvectors 
of $\sigma^\alpha_{~\beta}$ [as it is the case for the electric tide in the 
``silent universe" case \cite{bi:mps93},\cite{bi:silent}. 

The electric tidal tensor evolves according to 
\begin{eqnarray}
\nonumber 
{D {\cal E}^\alpha_{~\beta} \over D \tau} + {a' \over a} 
{\cal E}^\alpha_{~\beta} 
+ \vartheta {\cal E}^\alpha_{~\beta} + \delta^\alpha_{~\beta} 
\sigma^\gamma_{~\delta} {\cal E}^\delta_{~\gamma} - {3 \over 2} \biggl(
{\cal E}^\alpha_{~\gamma} \sigma^\gamma_{~\beta} + 
\sigma^\alpha_{~\gamma} {\cal E}^\gamma_{~\beta} \biggr) - 
\\ 
- {c^2 \over 2} \gamma_{\beta\eta} \biggl({\tilde \eta}^{\eta\gamma\delta} 
{\cal H}^\alpha_{~\gamma||\delta} + {\tilde \eta}^{\alpha\gamma\delta} 
{\cal H}^\eta_{~\gamma||\delta} \biggr) + 4\pi G a^2 \varrho_b(1+ \delta) 
\sigma^\alpha_{~\beta} = 0 \;,  
\end{eqnarray}
where we have rescaled the magnetic tide as ${\cal H}^\alpha_{~\beta} \equiv 
a^2 H^\alpha_{~\beta}$ and redefined the Levi--Civita tensor so that 
$\eta^{\alpha\beta\gamma} = \gamma^{-1/2} \epsilon^{\alpha\beta\gamma}$
(for simplicity we used the same symbol after rescaling). 

Finally, the magnetic Weyl tensor evolves according to
\begin{eqnarray}
\nonumber 
{D {\cal H}^\alpha_{~\beta} \over D \tau} + {a' \over a} 
{\cal H}^\alpha_{~\beta} 
+ \vartheta {\cal H}^\alpha_{~\beta} + \delta^\alpha_{~\beta} 
\sigma^\gamma_{~\delta} {\cal H}^\delta_{~\gamma} - {3 \over 2} \biggl(
{\cal H}^\alpha_{~\gamma} \sigma^\gamma_{~\beta} + 
\sigma^\alpha_{~\gamma} {\cal H}^\gamma_{~\beta} \biggr) + 
\\ 
+ {1 \over 2} \gamma_{\beta\eta} \biggl(\eta^{\eta\gamma\delta} 
{\cal E}^\alpha_{~\gamma||\delta} + \eta^{\alpha\gamma\delta} 
{\cal E}^\eta_{~\gamma||\delta} \biggr) = 0 \;. 
\end{eqnarray}

In the fluid--flow approach, besides the evolution equations, 
one has to satisfy a number of constraint equations. 
One has: the momentum constraint, which we rewrite in the form 
\be
\sigma^\alpha_{~\beta||\alpha} = {2 \over 3} \vartheta_{,\beta} \;, 
\ee
the ${\cal H}$--$\sigma$ constraint (which we already used to 
define the magnetic tide in terms of derivatives of the spatial 
metric), 
\be
{\cal H}^\alpha_{~\beta} = {1 \over 2} \gamma_{\beta\mu} 
\biggl(\eta^{\mu\gamma\delta} 
\sigma^\alpha_{~\gamma||\delta} + \eta^{\alpha\gamma\delta} 
\sigma^\mu_{~\gamma||\delta} \biggr) \;, 
\ee
the ${\rm div} ~{\cal E}$ constraint, 
\be
{\cal E}^\alpha_{~\beta||\alpha} = - \gamma_{\beta\mu}\gamma_{\alpha\nu} 
\eta^{\mu\lambda\gamma}\sigma^\nu_{~\lambda} {\cal H}^\alpha_{~\gamma} 
+ {8 \pi G \over 3} a^2 \varrho_b \delta_{,\beta} \;,
\ee 
and the ${\rm div} ~{\cal H}$ constraint 
\be
c^2 {\cal H}^\alpha_{~\beta||\alpha} = \gamma_{\beta\mu}\gamma_{\alpha\nu} 
\eta^{\mu\lambda\gamma}\sigma^\nu_{~\lambda} {\cal E}^\alpha_{~\gamma} \;.
\ee

The non--vanishing of ${\rm div} ~{\cal H}$, leading to the non--commutation of 
shear and tide, for the generic case of irrotational dust, follows from the 
recent analysis in Ref.\cite{bi:lesamepre}, where it was shown that requiring 
${\rm div} ~{\cal H}=0$ implies that ${\cal H}$ itself vanishes. 

In the above equations one also needs to know the three--metric 
$\gamma_{\alpha\beta}$. This can be obtained from the 
evolution equation 
\be
{\gamma_{\alpha\beta}}' = 2 \gamma_{\alpha\gamma} \vartheta^\gamma_{~\beta} \;,
\ee
which is however only valid in Lagrangian coordinates. 
In order to completely fix the spatial dependence of the metric one also 
needs to specify the energy constraint (the trace of the Gauss--Codacci 
relations), which we rewrite in the form
\be 
c^2 \bigl( {\cal R} - 6 \kappa \bigr) = \sigma^\alpha_{~\beta}
\sigma^\beta_{~\alpha} - {2 \over 3} \vartheta^2 - 4 {a' \over a}
\vartheta + 16 \pi G a^2 \varrho_b \delta \;.
\ee

Although we will not use the fluid--flow approach in this paper it
is interesting to have the complete form of the equations, with the correct 
powers of $c^2$ included, in order to understand the Newtonian meaning of 
the electric and magnetic tide. \\

We are now ready to discuss the fluid--flow approach 
within the Newtonian approximation.  
We just have to discuss the order in our $1/c^2$ expansion at which the 
various tensors enter the equations above. 
It is immediately clear that the mass continuity equation, the 
Raychaudhuri equation and the shear evolution equation, 
where no explicit powers of $c$ appear, just keep their form, once the 
various tensors are replaced by their Newtonian counterparts. So, we have 
\be
\bar \delta' + \bigl(1 + {\bar \delta} \bigr) {\bar \vartheta} = 0 \;,
\ee
\be
{\bar \vartheta}' + {a' \over a} {\bar \vartheta} + {1 \over 3} 
{\bar \vartheta}^2 + {\bar \sigma}^\alpha_{~\beta} 
{\bar \sigma}^\beta_{~\alpha} + 4 \pi G a^2 \varrho_b {\bar \delta} = 0 \;,
\ee
and 
\be 
{\bar \sigma}^{\alpha~'}_{~\beta} + {a' \over a} {\bar \sigma}^\alpha_{~\beta}
+ {2 \over 3} {\bar \vartheta} {\bar \sigma}^\alpha_{~\beta} + 
{\bar \sigma}^\alpha_{~\gamma} {\bar \sigma}^\gamma_{~\beta} - {1 \over 3} 
\delta^\alpha_{~\beta}
{\bar \sigma}^\gamma_{~\delta} {\bar \sigma}^\delta_{~\gamma} + 
{\bar {\cal E}}^\alpha_{~\beta} = 0 \;. 
\ee

On the other hand, by its very definition, the electric tide contains 
a contribution coming from the PN terms $\chi$ and $\pi^\alpha_{~\beta}$ 
because of the spatial curvature terms. 
It is however immediate to realize that, once the expressions of Section 4
for the PN tensors $\chi$ and $\pi^\alpha_{~\beta}$ are used, one recovers 
the simpler form, 
\be
{\bar {\cal E}}^\alpha_{~\beta} = \biggl( \nabla^\alpha \nabla_\beta -
{1 \over 3} \delta^\alpha_{~\beta} \nabla^2 \biggr) \varphi_g^{(L)} \;,
\ee
which, in Eulerian coordinates reduces to the standard 
form 
\be
{\bar {\cal E}}_{AB} = \varphi_{g,AB} - \frac{1}{3} \delta_{AB} \nabla_x^2 
\varphi_g \;. 
\ee

On the other hand, if we replace in Eq.(91), the Newtonian peculiar 
velocity--gradient tensor, 
we obtain the well--known result (e.g. Ref.\cite{bi:ellisvar}) that 
the magnetic tensor 
identically vanishes in the Newtonian limit. This can be very easily 
shown by either applying the formalism of Section 3, i.e. writing 
$\bar \vartheta^\alpha_{~\beta}$ through covariant derivatives of the 
Lagrangian velocity potential, or by writing the same tensor 
in terms of the Jacobian matrix of Section 3. The physics underlying this 
result is the conformal flatness of the Newtonian spatial sections, implying 
the commutation of spatial covariant derivatives. A simple consequence of this 
fact is that, 
at the Newtonian level, the ${\rm div} ~{\cal E}$ 
constraint reduces to 
\be
{\bar {\cal E}}^\alpha_{~\beta|\alpha} = {8 \pi G \over 3} a^2 \varrho_b 
{\bar \delta}_{,\beta} \;,
\ee
which, owing to our expression for ${\bar {\cal E}}$, turns out to be just the 
gradient of the Lagrangian Poisson equation, namely 
\be
\nabla^2 \varphi_g^{(L)} = 4 \pi G a^2 \varrho_b {\bar \delta} \;.
\ee

Let us now come to the tide evolution equation. 
In that evolution equation the circulation of the magnetic tensor is 
multiplied by 
$c^2$, which means that the PN part of ${\rm curl} ~{\cal H}$ is the source of
non--locality in the Newtonian electric tide evolution equation. 
On the other hand, if we look at the magnetic tide evolution equation, which 
starts to be non--trivial at the PN order, we see that 
${\rm curl} ~{\cal E}$ is consistently a PN quantity. 

The Newtonian meaning of the momentum constraint has been already 
discussed in Section 3. 
Also interesting is the ${\rm div} ~{\cal H}$ constraint, 
telling us that the general non--vanishing of the PN 
magnetic tensor (see also Ref.\cite{bi:lesamepre}), implies that the 
Newtonian shear and electric tide do not commute, i.e. they have different 
eigenvectors (viceversa, their non--alinement causes a non--zero 
${\rm div} ~{\cal H}$). Another possible 
version of this result is that the ratio of the velocity potential to 
the gravitational potential beyond the linear regime becomes 
space--dependent. 

To summarize, we can say that within the Newtonian approximation 
the fluid--flow approach in Lagrangian coordinates can be formulated in terms 
of mass continuity, Raychaudhuri and shear evolution equations plus 
the Newtonian ${\rm div} ~{\cal E}$ constraint, which closes the system, 
provided we remind the circulation--free character of the electric tide in this 
limit. Of course the direct use of a constraint to close the system of 
evolution equations, has the disadvantage of breaking the intrinsic 
hyperbolicity of the GR set of evolution equations, so that the entire 
method looses its basic feature. No ways out: this is the price to pay to the 
intrinsic non--causality of the Newtonian theory. 

The above discussion on the role of the PN magnetic tidal tensor, 
as causing non--locality in the Newtonian fluid--flow evolution equations, 
completely agrees with a similar analysis in Ref.\cite{bi:kp95}. 
The main difference being that the results reported here 
have been obtained directly in Lagrangian 
space, while they worked in non--comoving (i.e. Eulerian) coordinates. 
A different point of view on the subject is expressed by 
Bertschinger and Hamilton \cite{bi:bh94}, according to which 
the magnetic part of the 
Weyl tensor is non--vanishing already at the Newtonian level. According to 
Ref.\cite{bi:kp95}, the difference might be ``semantic"; what is 
most important is that there is general agreement on the 
fundamental fact that the Newtonian tide evolution is affected by 
non--local terms. 

Finally, let me mention how the equations for the ``silent universe" model 
\cite{bi:mps93},\cite{bi:mpsprl},\cite{bi:mps94},\cite{bi:silent} are 
obtained in this frame. If in all the equations above we were allowed to set 
${\cal H}^\alpha_{~\beta}$ identically to zero, then all the spatial gradient 
terms would disappear from the evolution equations. In that case each fluid 
element evolves independently of the others, apart from the self--consistency 
of the initial data, and the whole dynamical problem is reduced to the solution 
of a set of nine (because the shear, the electric tide and the metric can be 
shown to be aligned \cite{bi:barnes}) ordinary differential equations. 
The origin of this simplification is easy to understand from the physical 
point of view: setting the magnetic tide to zero forbids the 
presence of gravitational waves, which, given the absence of sound waves 
(because of the vanishing pressure), are the only way for our relativistic 
system to transport information among different fluid elements, during their 
evolution. 
However, as one could show by analysing the constraint equations 
above, the vanishing of the magnetic Weyl tensor requires very special 
initial conditions, which are incompatible with generic initial 
data for cosmological structure formation \cite{bi:mpsprl},\cite{bi:mps94}. 
On the other hand, on scales much larger than the horizon, the absence of causal 
correlation is precisely what we would expect to occur, so that the silent 
universe picture can be reasonably applied to describe the non--linear 
evolution of generic initial conditions on super--horizon scales
(see Refs.\cite{bi:mpsprl},\cite{bi:mps94}). In particular one can apply this 
model to study the asymptotic behaviour of large (super--horizon--size) 
patches of the universe in the presence of a non--zero cosmological constant; 
Bruni, Matarrese and Pantano \cite{bi:nohair} showed that in such a 
model most of the universe volume expands in a de Sitter--like fashion, 
consistently with the Cosmic No--hair Theorem. Contracting fluid elements, 
which exist even in the presence of a cosmological constant, 
generically collapse to a triaxial spindle singularity (see also 
Ref.\cite{bi:silent}). 

\section{Conclusions}

In these notes we have reviewed a Lagrangian approach to the evolution
of an irrotational and collisionless fluid in general relativity. 
The method is based on a standard $1/c$ expansion of the Einstein and 
continuity equations which leads to a purely Lagrangian derivation of 
the Newtonian approximation. One of the most important results in this 
respect is a simple expression for the Lagrangian metric in the Newtonian 
limit; this can be written in terms of the displacement vector 
${\bf S}({\bf q}, \tau) = {\bf x}({\bf q},\tau)  - {\bf q}$, from the 
Lagrangian coordinate ${\bf q}$ to the Eulerian one ${\bf x}$ of 
each fluid element, namely
\be
d s^2 = a^2(\tau) \biggl[ - c^2 d \tau^2 + \delta_{AB} 
\biggl(\delta^A_{~\alpha} + {\partial S^A({\bf q}, \tau) 
 \over \partial q^\alpha} \biggr) 
\biggl(\delta^B_{~\beta} + {\partial S^B({\bf q}, \tau) 
\over \partial q^\beta} \biggr) \biggr] \;.
\ee
The spatial metric is that of Euclidean space in time--dependent 
curvilinear coordinates, consistently with the intuitive notion 
of Lagrangian picture in the Newtonian limit. 

Next, we considered the post--Newtonian corrections to the metric and 
wrote equations for them. In particular, a simple and general equation for 
gravitational--wave emission from non--linear structures described through 
Newtonian gravity was given in Lagrangian coordinates, which is has the 
relevant feature of being non--local. 
A simple way to deal with the problem is to transform the equation
in Eulerian form, where it is easier to deal with the Laplacian operator
$\nabla_x^2$ (which has there the standard Euclidean form), 
obtain the Eulerian gravitational--wave tensor $\pi_{AB}$ and then go back 
to the Lagrangian expression through 
$\pi_{\alpha\beta} ({\bf q}, \tau) = {\bar {\cal J}}^A_{~~\alpha} 
{\bar {\cal J}}^B_{~~\beta} \pi_{AB}({\bf x}({\bf q},\tau), \tau)$. 
One obtains the Eulerian expression
\be
\nabla^2_x \pi_{AB} = \Psi^{(E)}_{v,AB} + \delta_{AB} \nabla_x^2 \Psi_v^{(E)}
+ 2 \biggl( \bar \vartheta \bar \vartheta_{AB} - 
\bar \vartheta_{AC} 
\bar \vartheta^C_{~~B} \biggr) \;, 
\ee
with $\nabla_x^2 \Psi_v^{(E)} = - \frac{1}{2} ( \bar \vartheta^2 - 
\bar \vartheta^A_{~B}  \bar \vartheta^B_{~A} )$. 
This formula allow to calculate the 
amplitude of the gravitational--wave modes in terms of the velocity 
potential $\Phi_v$, which in turn can be deduced from observational data on 
radial peculiar velocities of galaxies, applying the POTENT 
technique \cite{bi:potent}. 

In the standard case, where the cosmological perturbations form 
a homogeneous and isotropic random field, we can obtain a heuristic 
perturbative estimate of their amplitude in terms of the 
{\em rms} density contrast and of the ratio of the typical perturbation scale 
$\lambda$ to the Hubble radius $r_H=c H^{-1}$. One simply has
\be
{\pi_{rms} \over c^2} \sim 
\delta_{rms}^2 \biggl( {\lambda \over r_H} \biggr)^2 \;,
\ee
as it can be easily deduced from Eq.(80), specialized to an Einstein--de 
Sitter model. This effect gives rise to a stochastic background of 
gravitational waves which gets a non--negligible amplitude in 
the so--called {\em extremely--low--frequency} band, 
around $10^{-14}$ --  $10^{-15}$ Hz. 
On much smaller scales, where the effect might be even more relevant, 
pressure gradients and viscosity cannot be disregarded anymore and the 
entire formalism needs to be largely modified. 
One can roughly estimate that the present--day closure density of this
gravitational--wave background is 
\be
\Omega_{gw}(\lambda) \sim \delta_{rms}^4 
\biggl( {\lambda \over r_H} \biggr)^2 \;.
\ee 
In standard scenarios for the formation of structure in the universe, 
the typical density contrast on scales 
$1$ -- $10$ Mpc implies that $\Omega_{gw}$ is about $10^{-5}$ -- 
$10^{-6}$. 
One might speculate that such a background would give rise to 
secondary Cosmic Microwave Background anisotropies on intermediate angular 
scales through a sort of {\em tensor Rees--Sciama effect} (Matarrese and
Mollerach, in preparation). Let us finally stress that 
the amplitude of this gravitational--wave contribution, 
$\sim \delta^2 (\lambda /r_H)^2$, is an important counter--example to the rule 
given in the Introduction, according to which relativistic effects should be 
proportional to $\varphi_g/c^2 \sim \delta (\lambda / r_H)^2$. This is, 
in my opinion, an encouraging result, which should show to the skeptical 
reader that the application of relativistic cosmology to small scales can 
be much more than an elegant exercise. 

\section*{Acknowledgments.} I wish to thank my collaborator David Terranova 
for his essential contribution to the research project upon which these 
notes are based. The Italian MURST is acknowledged for partial 
financial support. Silvio Bonometto, Joel Primack and Antonello Provenzale are 
warmly thanked for having been so patient in waiting for my lecture notes to 
be completed.

\end{document}